\documentclass[superscriptaddress,aps,prb,twocolumn]{revtex4-2}
\usepackage{amsmath,amssymb,graphicx,dcolumn,bm,hyperref}
\usepackage{soul}
\usepackage{color}

\begin{document}

\title{Spin Nernst Effect of Antiferromagnetic Magnons in the Presence of Spin Diffusion}

\author{Hantao Zhang}
	\affiliation{Department of Electrical and Computer Engineering, University of California, Riverside, CA 92521, USA}	
\author{Ran Cheng}
	\affiliation{Department of Electrical and Computer Engineering, University of California, Riverside, CA 92521, USA}
	\affiliation{Department of Physics and Astronomy, University of California, Riverside, CA 92521, USA}

\begin{abstract}
Magnon spin Nernst effect was recently proposed as an intrinsic effect in antiferromagnets, where spin diffusion and boundary spin transmission have been ignored. However, diffusion processes are essential to convert a bulk spin current into boundary spin accumulation, which determines the spin injection rate into detectors through imperfect transmission. We formulate a diffusive theory to describe the detection of magnon spin Nernst effect with boundary conditions reflecting real device geometry. Thanks to the spin diffusion effect, the output signals in both electronic and optical detection grow rapidly with an increasing system size in the transverse dimension, which eventually saturate. Counterintuitively, the measurable signals are even functions of magnetic field, yielding optical detection more favorable than electronic detection.
\end{abstract}

\maketitle
\section{Introduction}
With the rapid growth of nano-electronics, it becomes increasingly demanding to develop energy-efficient means to process and transmit information. Magnons---the quanta of spin wave excitations---are promising alternative to electrons because they are charge neutral and can transport spin angular momenta in insulating materials without incurring Joule heating~\cite{Maekawa2017}. While ferromagnetic (FM) magnons exhibit fixed spin polarization determined solely by the magnetization, antiferromagnetic (AFM) magnons can carry both spin-up and spin-down polarizations similar to electrons~\cite{keffer1952theory,keffer1953spin,rezende2016diffusive,chen2016temperature}, forming an internal degree of freedom capable of encoding binary information. This unique property, along with the ultrafast spin dynamics, insensitivity to magnetic disturbance, have fertilized AFM magnonics as an emerging frontier in materials sciences~\cite{baltz2018antiferromagnetic}.

Concerning pure spin transport, the similarity between AFM magnons and electrons enables the magnonic analog of the SHE, known as the magnon spin Nernst effect (SNE)~\cite{cheng2016spin,zyuzin2016magnon,*kovalev2016spin,shiomi2017experimental,zhang2018spin}. In a thin-film geometry, the magnon SNE manifests as the generation of a transverse magnon spin current devoid of thermal Hall current by a longitudinal temperature gradient, where the spin polarization is perpendicular to the plane. To detect the electronic SHE, it is imperative to consider the spin diffusion process that converts a spin current into boundary spin accumulations~\cite{zhang2000spin}. This is because in real experiments, a bulk spin current is not directly measurable; only the boundary spin accumulation can produce detectable signals. In contrast, spin diffusion and boundary effects have not been considered so far in the magnon SNE~\cite{cheng2016spin,zyuzin2016magnon,*kovalev2016spin}. Consequently, we are not even able to ask for boundary spin accumulations within this intrinsic picture. A non-diffusive description also fails to capture the imperfect transmission of spin angular momenta between AFM magnons and metallic contacts, which is understandably essential to the electronic detection of magnon SNE. Moreover, unlike the magnon thermal Hall effect~\cite{matsumoto2014thermal,kim2016realization,ruckriegel2018bulk}, the magnon SNE is not accompanied by chiral edge currents, so boundary spin accumulations solely arise from bulk spin currents through the diffusion process. Therefore, a diffusive description of magnons is indispensable to build a correct understanding of the magnon SNE in AFM materials, and more importantly, to make meaningful predictions that can be compared with experiments.

In this paper, we formulate a diffusive theory to describe how the magnon SNE can be detected in the presence of spin diffusion and realistic boundary conditions in a thin-film AFM insulator. We first consider a prototype device geometry illustrated in Fig.~\ref{fig:model}, where the transverse boundaries are directly contacted to metallic leads. Driven by an applied in-plane temperature gradient, magnons with opposite spins diffuse towards opposite transverse boundaries and inject pure spin currents into the leads. The injected spins are subsequently converted into a charge voltage by the inverse SHE, producing an actual measurable signal. We find that the voltage output grows appreciably with an increasing system width until it eventually saturates. We then consider an isolated AFM insulator amenable to optical detection as illustrated in Fig.~\ref{fig:optical_potential}, where magnons accumulate on dead ends without injecting into leads. For both types of device geometry, we find that the detectable signals are even functions of the applied magnetic field along the N\'{e}el order (orthogonal to the plane), where the collinear ground state is well preserved below the spin-flop threshold. This is in sharp contrast to what one would naïvely obtain from a non-diffusive description. Consequently, it becomes fundamentally difficult to separate the inverse SHE voltage from the ubiquitous thermoelectric signal, yielding optical detection more favorable over electronic detection.

\section{Magnon spin diffusion}
Without loss of generality, let us consider a magnetic thin-film consisting of layered van der Waals AFM (such as MnPS$_3$ \cite{wildes1998spin} and FePS$_3$~\cite{lee2016ising}) with collinear N\'{e}el order perpendicular to the plane. Because the transport in the thickness dimension is suppressed, the system under consideration can be viewed as effectively two-dimensional. Because of strong easy-axis magnetic anisotropy and large spin magnitude, the long-range collinear ordering in such a system is well preserved even down to the monolayer limit~\cite{lee2016ising}. Therefore, we can ignore the quantum fluctuations of the N\'{e}el ground state and adopt a semi-classical picture of magnon excitations, which has been a widely accepted theoretical framework in studying magnonic transport~\cite{baltz2018antiferromagnetic,rezende2019introduction}. With a honeycomb lattice in mind, we attribute the existence of magnon SNE to the second-nearest neighboring Dzyaloshinskii-Moriya interaction which plays the role of an effective spin-orbit coupling for magnons~\cite{cheng2016spin,zyuzin2016magnon,bazazzadeh2021symmetry}. However, within the linear spin-wave regime, the special symmetry of the system does not break the rotational symmetry around the plane normal in the presence of the considered Dzyaloshinskii-Moriya interaction, which ensures that the $z$-component of the magnon spin is conserved~\cite{cheng2016spin,zyuzin2016magnon}, allowing us to separately define spin-up and spin-down magnons with respect to the plane normal (\textit{i.e.}, the $z$ axis). This property remains valid even when a magnetic field is applied along the $z$ direction so long as its strength is below the spin-flop threshold. With all these arguments being provided, however, the microscopic detail of the magnon SNE does not concern us here because our focus is the diffusion effect that governs the continuum limit of magnon transport.

\begin{figure}[t]
	\centering
  	\includegraphics[width=\linewidth]{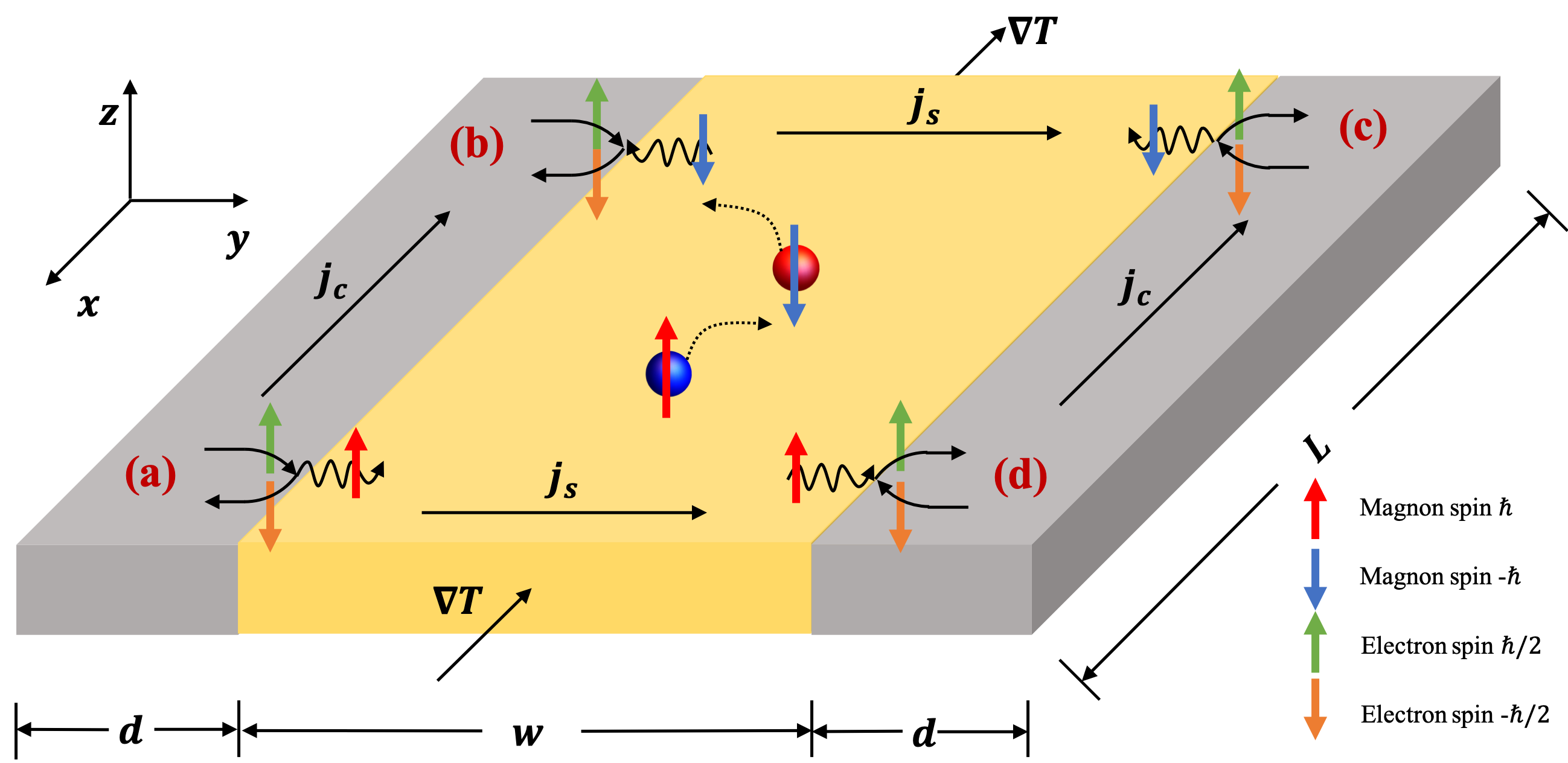}
	\caption{Illustration of system geometry and spin transmission processes at boundaries. A temperature gradient $\nabla T$ generates a transverse pure spin current $\bm{j}_{s}$ through the SNE in the AFM (yellow region). The spin current injects into the metallic leads (grey region) on both sides through four different spin transmission processes depicted by Feynman diagrams (a)--(d). The injected spin currents are converted into detectable voltages along $x$ through the inverse SHE. The system length along $x$ is $L$; the AFM width and the lead width are $w$ and $d$, respectively.}
	\label{fig:model}
\end{figure}

As illustrated in Fig.~\ref{fig:model}, a temperature gradient $\nabla T$ applied along the $x$ direction generates a pure spin current $j_s=-\sigma_s\nabla T$ in the $y$ direction, where the SNE coefficient $\sigma_s$ is a bulk quantity independent of boundaries. To solve the magnon spin diffusion in the $y$ direction, we make the following assumptions: 1) magnons of different spin species relax to the environment (lattice) individually without spin-flip scattering between magnons; 2) the momentum relaxation ascribing to spin-preserving processes is orders of magnitude faster than the spin relaxation due to spin-non-preserving processes. The first assumption is guaranteed by the aforementioned conservation of the $z$-component of spin, and similar assumption is made in Ref.~\cite{rezende2016diffusive,rezende2016theory}. The second assumption is generally true in clean and long-range ordered magnetic systems~\cite{xing2019magnon}. As a result, we can decouple the spin diffusion equations for each spin species, equating the magnon temperature to the environmental temperature locally, and only keep the $\partial_x T$ component which is fixed externally. In the natural units ($\hbar=k_B=e=1$), the current density of spin-up magnons is
\begin{align}
\label{eq:num_current_full}
j_y^{\uparrow} = -\sigma_{\uparrow} \partial_x T - D_{\uparrow} \partial_y \rho_{\uparrow}
\end{align}
where $\rho_{\uparrow}$ is the non-equilibrium density of spin-up magnons, $D_{\uparrow}$ is the magnon diffusivity, and $\sigma_{\uparrow}$ is the Nernst coefficient for spin-up magnons. Unless otherwise stated, all densities and currents below refer to the non-equilibrium contributions as there is no transport effect at thermal equilibrium. The equation for $j_y^{\downarrow}$ is similar and the total spin current density $j_s=j_{\uparrow}-j_{\downarrow}$ and $\sigma_s=\sigma_{\uparrow}-\sigma_{\downarrow}$. In the absence of magnetic fields, $\sigma_{\uparrow}=-\sigma_{\downarrow}$, $D_{\uparrow}=D_{\downarrow}$, and $\rho_{\uparrow}=\rho_{\downarrow}$ guaranteed by symmetry. A magnetic field perpendicular to the plane can break this symmetry, which will be discussed in the following.

Since we have ignored the spin-flip scattering between magnons with opposite spins, the spin continuity equation is respected separately by each spin species 
\begin{equation} \label{eq:continuity}
\frac{\partial \rho_{\gamma}}{\partial t} + \nabla \cdot \bm{j}^\gamma = -\frac{\rho_{\gamma}}{\tau_{\gamma}}, 
\end{equation}
where $\gamma=\uparrow, \downarrow$ and $\tau_{\gamma}$ is the effective spin-relaxation time due to spin-non-conserving scattering with phonons, impurities, etc. We treat $\tau_{\gamma}$ as a phenomenological parameter; $\tau_{\uparrow}=\tau_{\downarrow}$ in the absence of magnetic fields. We will focus on the linear response regime such that the spatial inhomogeneity of transport coefficients in Eq.~\eqref{eq:num_current_full} is negligible, which requires $|L\partial_xT/T|\ll1$ with $L$ the length in the $x$ direction (see Fig.~\ref{fig:model}). At steady state, $\partial \rho_{\gamma}/\partial t = 0$, then inserting Eq.~\eqref{eq:num_current_full} into Eq.~\eqref{eq:continuity} gives $\nabla^2\rho_{\gamma}=\rho_{\gamma}/\lambda_{\gamma}^2$, where $\lambda_{\gamma}=\sqrt{D_\gamma\tau_\gamma}$ is the effective spin diffusion length for spin $\gamma$. When the system width $w$ far exceeds its length $L$, the diffusion process becomes effectively a one dimensional problem. Therefore,
\begin{eqnarray}
\label{eq:diffusion_eq}
\frac{\partial^{2} \rho_{\gamma}}{\partial y^{2}} = \frac{\rho_{\gamma}}{\lambda_{\gamma}^{2}},
\end{eqnarray}
which requires two boundary conditions to solve for each $\gamma$. In the right (left) leads, the spin density of electrons $\rho_R$ ($\rho_L$) satisfies the same diffusion equation governed by a different spin diffusion length $\lambda_e=\sqrt{D_e\tau_e}$ where $D_e$ is the spin diffusivity and $\tau_e$ is the spin relaxation time of electrons.

Solving Eq.~\eqref{eq:diffusion_eq} together with the spin diffusion equations for $\rho_{R/L}$ calls for proper boundary conditions determined by the interfacial spin transmission between magnons and electrons. As illustrated in Fig.~\ref{fig:model}, there exists four different scattering processes depending on the spin polarization and flow direction of magnons, each involving an electron that releases an angular momentum of either $\hbar$ or $-\hbar$ through spin-flip and a magnon that balances the change of the electron spin. Specifically, magnons are emitted by electrons in process (a) and process (c), while they are absorbed by electrons in (b) and (d). In the absence of magnetic fields, (a) and (b) form a time-reversal pair, which should take place with the same rate, so do (c) and (d).
These four interfacial processes are characterized by four conductance parameters that can be calculated by extending the linear response theory previously formulated in ferromagnets~\cite{zhang2012spin,chen2016temperature}. Assuming identical interfacial properties and detailed balance, however, we can reduce these four parameters into two such that (a) and (d) [(b) and (c)] are represented by the same conductance $G_{\uparrow}$ ($G_{\downarrow}$)~\cite{troncoso2020spin}, which is proportional to the interfacial spin-mixing conductance (see Appendix). For example, on the right interface ($y=w/2$), the Ohm's law associated with (c) and (d) determines two spin current densities crossing the interface: $j^{\uparrow}_{\rm int}=G_{\uparrow}[\rho_{\uparrow}(w/2)-\rho_R(w/2)]$ and $j^{\downarrow}_{\rm int}=G_{\downarrow}[\rho_{\downarrow}(w/2)-(-\rho_R(w/2))]$. Spin continuity further requires that $j^{\uparrow}_{\rm int}=j_y^{\uparrow}(w/2)$ and $j^{\downarrow}_{\rm int}=j_y^{\downarrow}(w/2)$, and that $j^{\uparrow}_{\rm int}-j^{\downarrow}_{\rm int}=-D_e\left.\frac{\partial\rho_R}{\partial y}\right|_{w/2}$. These relations constitute three independent conditions at $y=w/2$. Similarly, there are three boundary conditions associated with processes (a) and (b) on the opposite interface at $y=-w/2$. At the dead ends $y=\pm(w/2+d)$ with $d$ the lead width, spin currents must vanish identically, which provides another two independent conditions. Including everything, we finally obtain a set of eight boundary conditions grouped into four relations as
\begin{subequations} \label{eq:boundary_condition}
\begin{align}
&G_{\gamma} \left[ \rho_{L} \mp  \rho_{\gamma} \left(-\frac{w}2\right) \right] = \pm j^{\gamma}_{y} \left(-\frac{w}2\right),  \label{eq:bc_left} \\
&G_{\gamma} \left[ \rho_{R} \mp  \rho_{\gamma} \left(\frac{w}2\right) \right] = \mp j^{\gamma}_{y} \left(\frac{w}2\right),  \label{eq:bc_right} \\
&-D_{e} \left. \frac{\partial \rho_{R/L}}{\partial y} \right|_{\pm w/2} =  j^{\uparrow}_{y}\left(\pm\frac{w}2\right) -  j^{\downarrow}_{y}\left(\pm\frac{w}2\right), \label{eq:bc_both} \\
&-D_{e} \left. \frac{\partial \rho_{R/L}}{\partial y}\right|_{\pm (w/2 + d)} = 0, \label{eq:bc_exterior}
\end{align}
\end{subequations}
where $\gamma = \uparrow$ ($\downarrow$) in Eq.~\ref{eq:bc_left} and~\ref{eq:bc_right} and $R$ ($L$) in Eq.~\eqref{eq:bc_both} and~\eqref{eq:bc_exterior} are linked to the upper (lower) sign of $\pm$ and $\mp$ appearing in these equations. It should be noted that $G_{\gamma}$ is the same on both boundaries and independent of $\rho_{\gamma}$ and $\rho_{R/L}$, which is true in the linear response regime for identical leads on both sides. Invoking the above boundary conditions on Eq.~\eqref{eq:diffusion_eq}, we are able to solve the magnon density $\rho_{\uparrow}(y)$ and $\rho_{\downarrow}(y)$ for $-w/2<y<w/2$ and the electron spin density $\rho_L(y)$ and $\rho_R(y)$ for $-(w/2+d)<y<-w/2$ and $w/2<y<w/2+d$, respectively.

\section{Electronic Detection}
\begin{figure}[t]
	\centering
  	\includegraphics[width=0.95\linewidth]{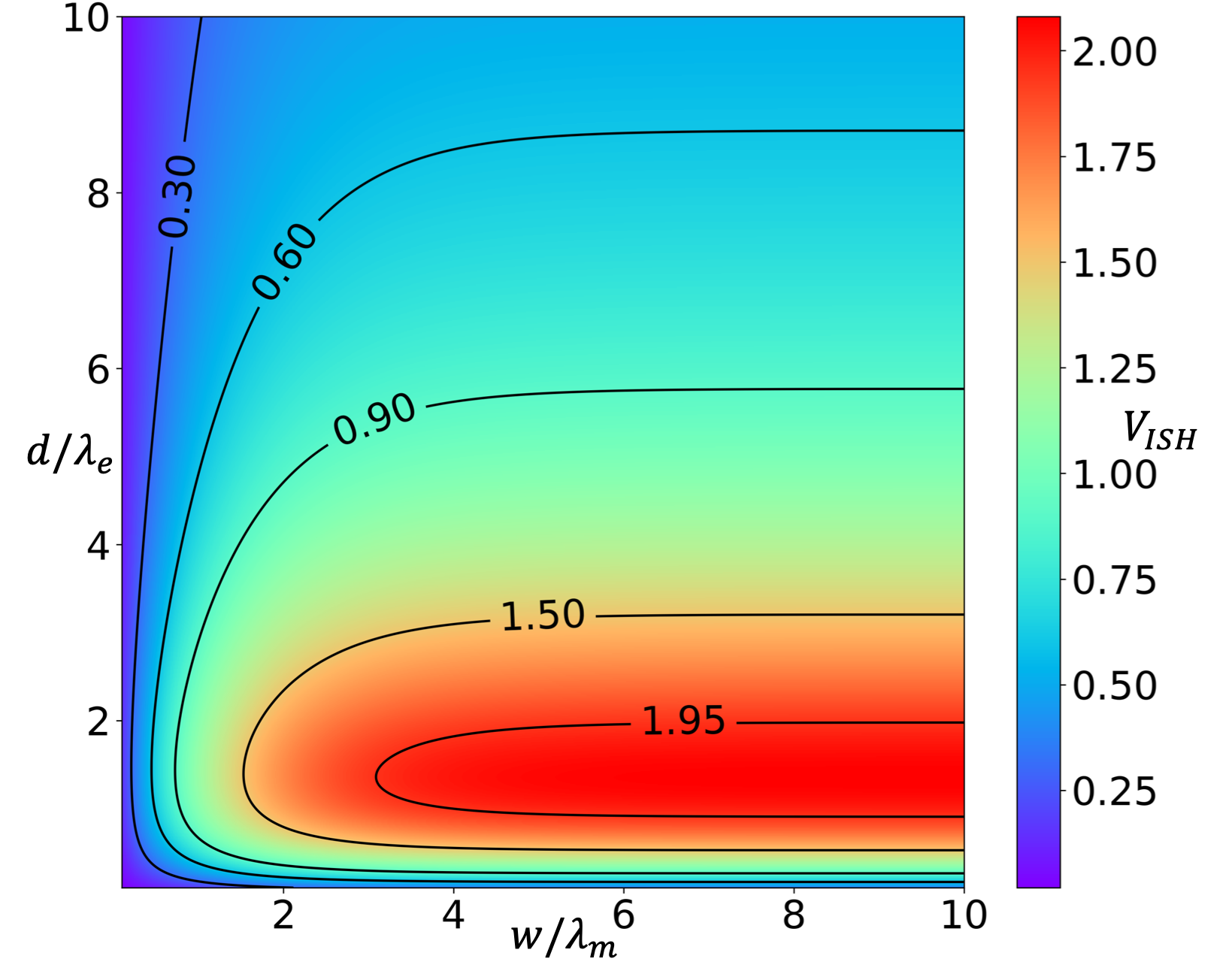}
	\caption{Inverse SHE voltage resulting from the spin currents injected into the leads as a function of $w/\lambda_{m}$ and $d/\lambda_{e}$ for $\eta_m=\eta_e=16$, which is estimated by using materials parameters in MnPS$_3$ (see the Appendix). Unit: $-\partial_xT(\sigma\theta_sL/eG\lambda_e)$.}
	\label{fig:ISH_V}
\end{figure}

In the leads, the inverse SHE converts the injected spin angular momenta into an electrical current $\bm{j}_{L/R}= \sigma_{c} [\bm{E}_{L/R} - \theta_s/(eg_F) \hat{z}\times\bm{\nabla} \rho_{L/R}]$~\cite{cheng2016dynamic}, where $\sigma_{c}$ is the conductivity of the leads, $\bm{E}_{L/R}$ is the electric field, and $\theta_s$ is the spin Hall angle, and $g_F$ is the density of states at the Fermi level. With the open boundary condition $\bm{j}_{L/R} = 0$, the inverse SHE generates a voltage $V_{L/R}=L/d\int dy E_{L/R}$ along the $x$ direction. If the two leads are made of identical materials and the interfacial properties are the same on both sides, symmetry guarantees that $V_L=V_R=V$ under the SHE geometry. If the two leads and the corresponding interfaces are different, $V_{L}$ and $V_{R}$ are likely to be different, which might be a useful strategy for measurement. In the absence of magnetic fields, time-reversal symmetry guarantees that $\sigma=\sigma_{\uparrow}=-\sigma_{\downarrow}$, $\lambda_m=\lambda_{\uparrow}=\lambda_{\downarrow}$, $D_m=D_{\uparrow}=D_{\downarrow}$, and $G=G_{\uparrow}=G_{\downarrow}$. Retrieving all physical constants from the natural units, we obtain
\begin{eqnarray} \label{eq:ISH_V_no_field}
&&V = -\partial_x T\frac{\sigma\theta_{s} L}{eGd} \frac{\eta_m\eta_e\tanh\frac{d}{2\lambda_e}}{\eta_m+\coth\frac{w}{2\lambda_m}\left(1+2\eta_e\coth\frac{d}{\lambda_e}\right)}, 
\end{eqnarray}
where $e$ is the electron charge, $\eta_m=G\lambda_m/D_m$ and $\eta_e=G\lambda_e/D_e$ are dimensionless parameters, and $w$, $d$, and $L$ describe the device geometry illustrated in Fig.~\ref{fig:model}. The estimation of $\eta_m$ and $\eta_n$ are discussed in the Appendix. The spin diffusion effect of magnons is reflected in the ratio $w/\lambda_m$, while $d/\lambda_e$ affecting the electron spin diffusion in the leads can be varied independently. In the ballistic limit that $w\ll\lambda_m$, the $\coth w/2\lambda_m$ factor in the denominator of Eq.~\eqref{eq:ISH_V_no_field} blows up, thus the actual output voltage is highly suppressed. To linear order in $w/\lambda_m$, we have $V=-\partial_xT(wL/2d)(\sigma\theta_{s}/eD_m)\eta_e\tanh (d/2\lambda_e)/[1+2\eta_e\coth (d/\lambda_e)]$, which is linear in $w$. On the other hand, when both magnons and electrons are in the diffusive limit, \textit{i.e.} $w\gg\lambda_m$ and $d\gg\lambda_e$, Eq.~\ref{eq:ISH_V_no_field} becomes
\begin{align}
    V=-\partial_xT\frac{\sigma\theta_{s} L}{ed}\frac{G \lambda_{e} \lambda_{m}}{D_eD_m +  G(D_e\lambda_m+2D_m\lambda_e)},
\end{align}
which is independent of $w$. Figure~\ref{fig:ISH_V} plots the output $V$ as a function of $w/\lambda_m$ and $d/\lambda_e$ for typical materials parameters of MnPS$_3$ (see the Appendix). The essential pattern of Fig.~\ref{fig:ISH_V} is preserved even by varying $\eta_m$ and $\eta_e$. While $V$ increases monotonically towards saturation with an increasing $w/\lambda_m$, it varies non-monotonically with $d/\lambda_e$, where the maximum appears for $d$ being comparable to $\lambda_e$. These features suggest that: 1) The diffusion effect of magnons can significantly facilitate the electronic detection of the magnon SNE. The saturation voltage output for $w\gg\lambda_m$ can be orders of magnitude larger than that in the ballistic limit. 2) The spin diffusion of electrons in the leads can either enhance or suppress the output, so the dimensions and the material properties of the leads should be optimized.

The fact that the output voltage in both leads are the same makes it difficult to separate the magnon SNE from thermoelectric effects. Therefore, it is natural to consider applying a magnetic field to imbalance different spin species. Here we only consider a field perpendicular to the plane with a strength below the spin-flop threshold, which breaks the degeneracy of the spin-up and spin-down magnons but cannot change the collinear ground state. However, a scrutiny over the spin transmission processes in Fig.~\ref{fig:model} suggests a negative conclusion. Suppose that the Zeeman interaction lowers the gap of spin-up magnon band, enlarging its population, whereas spin-down magnons experience the opposite change. In the linear response regime, the intensity of process (a) dubbed $I_a$ and that of process (d) dubbed $I_d$ will increase by exactly the same amount $\Delta I_+$. Similarly, $I_b$ and $I_c$ will decrease by the same amount $\Delta I_-$. Accordingly, the total amount of spin injection from the left lead, determined by $I_a+I_b$, will change by $\Delta I_+-\Delta I_-$. This is exactly the same as the change of spin injection into the right lead determined by $I_c+I_d$. Moreover, since (a) and (b) form a time-reversal pair, reversing the magnetic field direction will lead to a decrease of $\Delta I_-$ in $I_a$ and an increase of $\Delta I_+$ in $I_b$, leaving the overall change of $I_a+I_b$ still $\Delta I_+-\Delta I_-$ [same argument applies to (c) and (d) as well]. This means $V(B) = V(-B)$, namely, the voltage output is an even function of $B$. As a result, the output voltage $V$ must be quadratic in the magnetic field to the lowest order. If we only consider the magnon injection processes (b) and (d) while ignoring their emission partners (a) and (c), we could arrive at the wrong conclusion where there is a linear $B$ dependence. 

Now we justify the above conclusion by expanding the inverse SHE voltages $V_L$ and $V_R$ arising from the two leads with respect to $B$. For weak fields, $\sigma_\gamma=\sigma\pm\sigma'B+\sigma''B^2/2$ where $\sigma'=\partial_B\sigma(0)$, $\sigma''=\partial_B^2\sigma(0)$, and the $+$ ($-$) sign corresponds to $\gamma=\uparrow$ ($\downarrow$). Similar expansions apply to all other parameters such as $\eta_m$ and $\eta_e$. The magnetic field dependencies of $\lambda_{m}$, $D_{m}$ and $G$ are included in the expansion of $\eta_{m}$ and $\eta_{e}$. Solving $V_L$ and $V_R$ from Eq.~\eqref{eq:diffusion_eq} under spin-specific boundary conditions (\textit{i.e.}, $\gamma=\uparrow$ or $\downarrow$ in Eq.~\eqref{eq:boundary_condition}), we find that the linear term of $B$ vanishes identically in both $V_L$ and $V_R$ while they share the same $B^2$ term. Therefore, we have $V_L=V_R=V(B)$ even in the presence of a magnetic field. Letting $\Delta V\equiv V(B)-V(0)$, we obtain
\begin{eqnarray} \label{eq:ISH_V_field}
 \frac{\Delta V}{V(0)} = \frac{B^{2}}{\eta_{m} \sigma \mathfrak{D}} \left( \frac{a }{1 + \eta_{m}} + b + c \right),
\end{eqnarray}
where $\mathfrak{D} = 1+ \eta_{m} + 2 \eta_{e}$, $a = -\eta_{m}^{\prime 2}\sigma (1 + 2\eta_{e}) + 2\eta_{m}^{\prime} (\eta_{m} \eta_{e}^{\prime} \sigma + \eta_{e} \sigma^{\prime})$, $b = \eta_{m}^{\prime \prime} \sigma/2 + \eta_{m}(1+\eta_{m})\sigma^{\prime \prime}/2 + \eta_{m}^{\prime}\sigma^{\prime}$, and $c = \eta_{m}^{\prime \prime} \eta_{e} \sigma - \eta_{m} \eta_{e}^{\prime \prime} \sigma + \eta_{m} \eta_{e} \sigma^{\prime \prime}$. In the expansions, we have ignored the very insensitive field dependence of $\theta_{s}$, $D_{e}$ and $\lambda_{e}$. Since a magnetic field cannot make $V_L$ and $V_R$ different, electronic detection of the magnon SNE turns out to be an unreliable approach.

\section{Optical Detection}
\begin{figure}[t]
	\centering
  	\includegraphics[width=\linewidth]{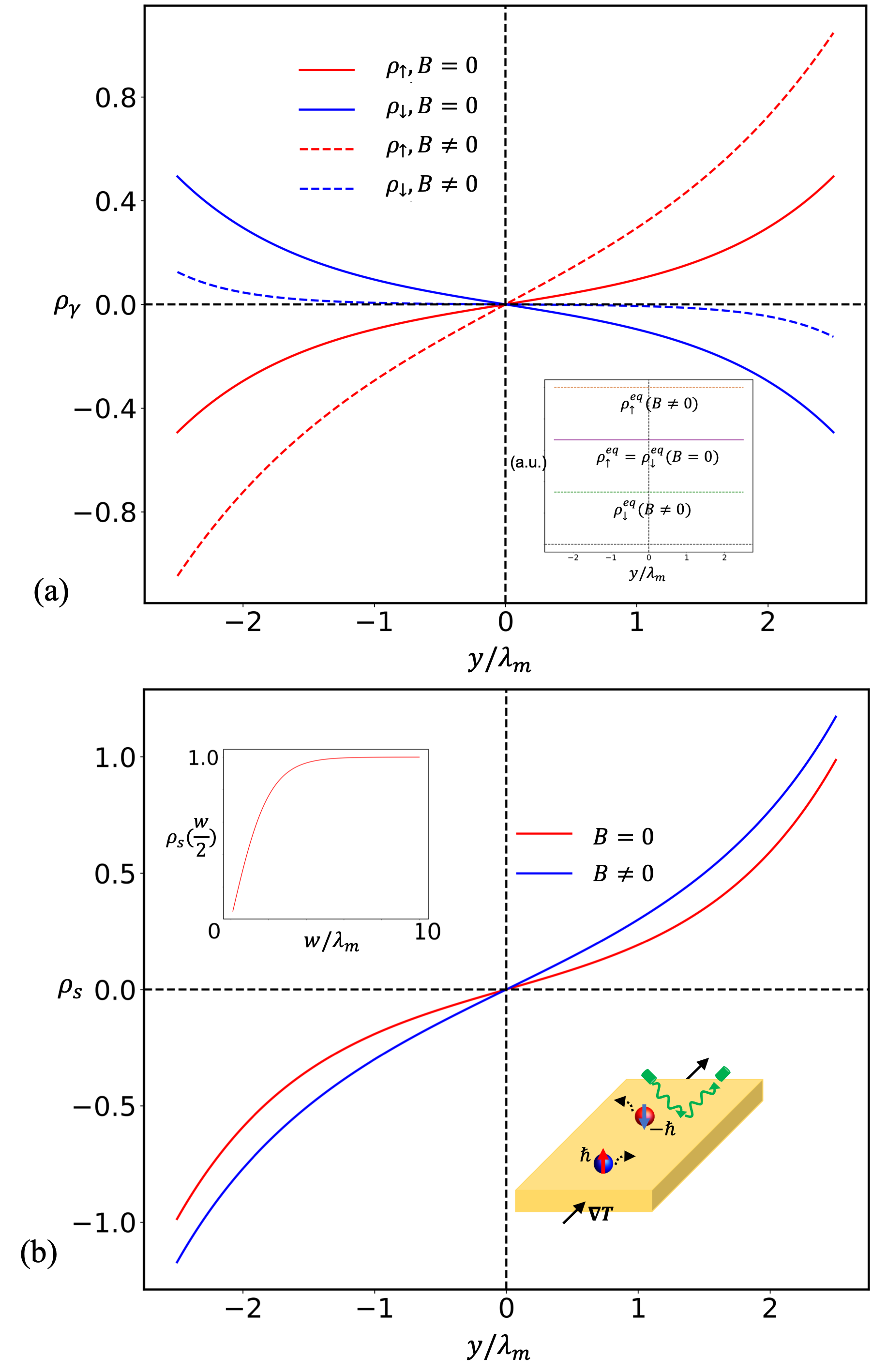}
	\caption{(a) Non-equilibrium density $\rho
	_{\uparrow, \downarrow}$. Inset: Equilibrium magnon density $\rho^{eq}_{\uparrow,\downarrow}$. (b) Non-equilibrium magnon spin density $\rho_s$ as a function of $y$ for open boundaries and $w=5\lambda_m$. Insets: Non-equilibrium spin density on the right edge $\rho_{s}(w/2)$ as a function of $w/\lambda_{m}$ (upper-left); illustration of device geometry (lower-right). Unit: $-2\partial_xT(\sigma\lambda_m/D_m)$.}
	\label{fig:optical_potential}
\end{figure}

The boundary spin accumulation arising from the diffusive magnon SNE can be detected optically without metallic leads. Different from electronic detection, an optical detection reacts to both equilibrium and non-equilibrium magnons. To the lowest order, the equilibrium magnon spin density, $\rho_{s}^{eq} = \rho_{\uparrow}^{eq}-\rho_{\downarrow}^{eq}$, is linear in a perpendicular magnetic field. But this part is not related to the magnon SNE because it does not diffuse and is independent of $y$ as shown in the inset of Fig.~\ref{fig:optical_potential}(a). So we need to focus on the non-equilibrium contribution. Accordingly, we use $j_{\gamma}(\pm w/2)=0$ as the boundary conditions and solve the non-equilibrium magnon density $\rho_{\uparrow}$ and $\rho_{\downarrow}$ from Eq.~\eqref{eq:diffusion_eq}, which is plotted in Fig.~\ref{fig:optical_potential}(a). The net non-equilibrium spin density $\rho_s=\rho_{\uparrow}- \rho_{\downarrow}$ is
\begin{equation} \label{eq:chemical_potential_optical}
\rho_{s}(y) = -2\partial_xT\frac{\sigma\lambda_m}{D_m} \text{sech} \frac{w}{2\lambda_m} \sinh \frac{y}{\lambda_m},
\end{equation}
which is plotted in Fig.~\ref{fig:optical_potential}(b). Because $\rho_s(y)=-\rho_s(-y)$ is an odd function, the profile can be unambiguously probed via spin-resolved magnon-photon interactions if the spatial resolution is higher than $1/w$. Typical optical measurements are not able to discern $ \rho_{\uparrow}(y)$ and $ \rho_{\downarrow}(y)$ individually; only the spin density $\rho_s(y)$ can be measured. On the edges, $\rho_s(\pm w/2)\sim\pm\tanh(w/2\lambda_m)$, which increases with an increasing ratio of $w/\lambda_m$ until it eventually saturates. This behavior, shown in the inset of Fig.~\ref{fig:optical_potential}(b), is similar to the case of electronic detection. 

Although thermoelectric effects are no longer a concern in optical detection, it is instructive to examine how $\rho_s$ depends on a perpendicular magnetic field. In the diffusive limit that $w \gg \lambda_{\gamma}$ ($\gamma=\uparrow,\downarrow$), we expand the spin-dependent quantities up to quadratic order in $B$. On the right edge, the change of non-equilibrium spin density $\Delta \rho_s = \rho_s(B, w/2) - \rho_s(0, w/2)$ is obtained as
\begin{eqnarray}
 \label{eq:optiacal_potential_field}
\frac{\Delta \rho_s}{\rho_s(0, w/2)} = B^{2} \frac{(\zeta \sigma)^{\prime \prime}}{2\zeta \sigma}
\end{eqnarray}
where $\zeta = \lambda_m/D_m$, and $\sigma$ are defined for $B=0$; $\zeta^{\prime \prime}=\partial_B^2\zeta(0)$ and $\sigma^{\prime \prime}=\partial_B^2\sigma(0)$. Similar to the electronic detection, the change of non-equilibrium magnon spin density on each boundary is an even function of $B$, so reversing the field direction leads to the same result. However, we are not able to determine the sign of $(\zeta\sigma)^{\prime\prime}/\zeta\sigma$ in Eq.~\eqref{eq:optiacal_potential_field}, so a perpendicular magnetic field can either enhance or suppresse the boundary spin accumulation depending on materials.

To better visualize the influence of magnetic field, we exaggerate the changes of $\zeta$ and $\sigma$ induced by $B$ in plotting the profile of $\rho_{\uparrow, \downarrow}(y)$ and $\rho_s(y)$ in Fig.~\ref{fig:optical_potential}. In fact, the quadratic $B$-dependence and the anti-symmetric spatial distribution of the spin density, is not unique to the magnon SNE in AFM. It was also observed in the electron SHE under optical detection~\cite{kato2004observation}.

\section{Materials Estimate}
Comparing with the electronic detection where the output voltage is difficult to be separated from thermoelectric effects even in the presence of magnetic field, optical detection turns out to be the preferred method to observe the diffusive magnon SNE thanks to the anti-symmetric profile of $ \rho_s(y)$.
Magneto-optic Kerr effect microscopy is a well-established optical detection~\cite{kato2004observation,van2014optical,stamm2017magneto}, but it does not provide enough spatial resolution to measure systems on or below the micrometer scale. Recent development in the nitrogen-vacancy (NV) center magnetometer~\cite{taylor2008high,maletinsky2012robust,van2015nanometre,du2017control,wang2020quantum}, on the other hand, exhibits remarkable sensitivity combined with a high spatial resolution, making it quite promising to measure the spatial profile of $\rho_s(y)$, hence the magnon SNE. 

While the magnon SNE was theoretically predicted and experimentally explored in MnPS$_{3}$~\cite{cheng2016spin,shiomi2017experimental}, recent studies suggest that its variance (such as CrSiTe$_{3}$) can exhibit a much larger SNE coefficient~\cite{bazazzadeh2021symmetry}. Therefore, we choose the largest reported value, $\sigma_{\uparrow}=\sigma_{\downarrow} \approx 2.5 \times 10^{-2} k_{B}/ \hbar$, for estimation. In the absence of magnetic fields, we take $\tau_m \approx 10$ns~\cite{shen2020magnon}, $\lambda_m \approx 1 \mu$m~\cite{xing2019magnon}, and $w\gg\lambda_m$. For a thin film consisting of about 10 layers~\cite{lin2016ultrathin}, a temperature gradient $\partial_x T$ on the order of $0.1 \sim 1$K/$\mu$m~\cite{walter2011seebeck,holanda2017spin,holanda2017longitudinal,kryder2008heat} will produce an areal spin density of roughly $10^{13}\sim10^{14}\hbar$/m$^2$ on the edge. Assuming a pixel size of 20nm $\times$ 20nm and a distance of 20nm between the NV center and the material surface~\cite{maletinsky2012robust}, we estimate that the edge spin accumulation of magnons will generate a static magnetic field acting on the NV center on the order of $1\sim10$nT, which is within the sensitivity~\cite{wang2020quantum}.

In summary, we have formulated a diffusive theory to describe the detection of magnon SNE in antiferromagnets with collinear N\'{e}el order, providing experimentally measurable predictions missed in previous theoretical studies. Owing to the magnetoelectric effects which mixes with the output voltage, optical detection turns out to be more reliable than electronic detection. The NV center magnetometer is able to fulfill this function. We anticipate that our findings can inspire ongoing experiments of the magnon SNE in AFM insulators.

\begin{acknowledgments}
We acknowledge inspiring discussions with X. Chen, C. Du, and A. Balandin. This work is supported by the Air Force Office of Scientific Research under grant FA9550-19-1-0307.
\end{acknowledgments}

\appendix*
\section{Estimation of $\eta_{m}$ and $\eta_{e}$}
The dimensionless parameters $\eta_m = G \lambda_m / D_m$ and $\eta_e = G \lambda_e / D_e$ both depend on the interfacial conductance $G$, which converts an interfacial spin density on one side to an spin current density on the other side of the interface. Calculating $G$ can be very sophisticated, but fortunately we only need a rough estimate of its magnitude. Basing on Ref.~\cite{rezende2016diffusive,chen2016temperature,zhang2020magnon,li2020spin}, we evaluate $G$ from the real part of the interfacial spin-mixing conductance $g_r$.

The contribution of a particular magnon mode, either spin up or spin down, with wave vector $\bm{k}$ to the interfacial spin current density is $j^{\gamma}_{\rm int}(\bm{k}) = g_{r} \varepsilon_{k}^\gamma \delta n_{k}^\gamma$ ($\gamma = \uparrow \text{or} \downarrow$), where $\varepsilon_k^\gamma$ and $\delta n_k^\gamma$ is the energy and the non-equilibrium distribution of magnons, respectively. In the linear response regime, $\delta n_{k}^\gamma = \mu_\gamma \partial n_k^\gamma / \partial \mu $, where $n_k^\gamma$ is the Bose-Einstein distribution function and $\mu_\gamma$ is the chemical potential. So, the total contribution from magnons with spin $\gamma$ is

\begin{eqnarray}
\label{eq:spin_current}
j^{\gamma}_{\rm int} = g_{r} \mathcal{V} \int \frac{d^3 k}{(2\pi)^3} \varepsilon_{k}^\gamma \left. \frac{\partial n_k^\gamma}{\partial \mu}\right|_{\mu = 0} \mu_\gamma
\end{eqnarray}
where $\mathcal{V}$ is the volume of primitive cell and 
\begin{align}
    \left. \frac{\partial n_k^\gamma}{\partial \mu}\right|_{\mu = 0} = \frac{1}{k_{B}T} \frac{ e^{\varepsilon_k^\gamma / k_{B}T}}{(e^{\varepsilon_k^\gamma / k_{B}T}-1)^{2}}.
\end{align}
Meanwhile, $\mu_\gamma$ can be related to the interfacial spin accumulation $\rho_\gamma$ as
\begin{eqnarray}
\label{eq:magnon_accumulation}
\rho_{\gamma} = \hbar \int \frac{d^3 k}{(2\pi)^3} \left. \frac{\partial n_k^\gamma}{\partial \mu}\right|_{\mu = 0} \mu_\gamma.
\end{eqnarray}
Therefore, the effective interfacial conductance $G$ can be obtained as
\begin{align}
G = \frac{j^{\gamma}_{\rm int}}{\rho_{\gamma}} = \left. \frac{g_r \mathcal{V}}{\hbar} \frac{ \int d^3 k \varepsilon_k (\partial n_k^\gamma / \partial \mu)  }{ \int d^3 k (\partial n_k^\gamma/\partial \mu)} \right|_{\mu = 0},  \label{eq:j_over_rho}
\end{align}
whose dimension is m/s in SI unit. In the absence of magnetic fields, $G$ is the same for both spin-up and spin-down bands. We ignore the dependence of $G$ on magnetic fields for weak fields.

Next, we use the material parameters of MnPS$_{3}$~\cite{wildes1998spin}, assume the second-nearest neighboring Dzyloshinskii-Moriya interaction to be $D_2=0.2$meV\cite{cheng2016spin}, and $g_{r} = 10^{18} \text{m}^{-2}$~\cite{chen2016temperature,zhang2020magnon} to calculate $G$ numerically using Eq.~\eqref{eq:j_over_rho}. By allowing temperature $T$ to vary between $10$K and $40$K, we find that $G$ varies from $526$m/s to $1645$m/s. Since $\lambda_{m} = \sqrt{D_{m} \tau_{m}}$ and $\lambda_{e} = \sqrt{D_{e} \tau_{e}}$, $\eta_{m}$ and $\eta_{e}$ can also be expressed as $\eta_{m} = G \tau_{m} / \lambda_{m}$ and $\eta_{e} = G \tau_{e} / \lambda_{e}$. Because we do not know the exact value of $\tau_m$ for MnPS$_{3}$, we take a typical estimate that $\tau_{m} \approx 10$ns~\cite{shen2020magnon}. In addition, we use $\lambda_{m} \approx 1\mu$m~\cite{xing2019magnon}, $\tau_{e} \approx 10$ps~\cite{fang2017determination} and $\lambda_{e} \approx 1$nm~\cite{ma2018spin}. Combining everything, we finally obtain $\eta_m \approx \eta_e \approx 16$ at $T=40$K.



\bibliography{manuscript_of_Diffusive_spin_Nernst_effect}

\end{document}